\documentclass[12pt]{article}
\usepackage{fancybox}
\usepackage[dvips]{graphicx}
\usepackage{latexsym}
\makeatletter
\@addtoreset{equation}{section}

\makeatother

\begin{document}
\begin{titlepage}
 
$\mbox{ }$
\begin{flushright}
\begin{tabular}{l}
KUNS-1776\\
hep-th/0204078\\
Apr 2002
\end{tabular}
\end{flushright}

~~\\
~~\\
~~\\

\vspace*{0cm}
    \begin{Large}
       \vspace{2cm}
       \begin{center}
         {Matrix model with manifest general coordinate invariance}      \\
       \end{center}
    \end{Large}

  \vspace{1cm}

\begin{center}
           Takehiro A{\sc zuma}$^{\dag}$\footnote{
e-mail address : azuma@gauge.scphys.kyoto-u.ac.jp}
and
           Hikaru K{\sc awai}$^{\dag \ddag}$\footnote
           {
e-mail address : hkawai@gauge.scphys.kyoto-u.ac.jp} \\
\vspace*{1cm}
       $^{\dag}${\it Department of Physics, Kyoto University,
Kyoto 606-8502, Japan}\\
\vspace*{1cm}
       $^{\ddag}$ {\it Theoretical Physics Laboratory,
      RIKEN (The Institute of Physical and Chemical Research)\\
       2-1 Hirosawa, Wako, Saitama 351-0198, Japan}
\end{center}

\vfill

\begin{abstract}
\noindent
We present a formulation of a matrix model which manifestly possesses
the general coordinate invariance when we identify the large $N$
matrices with differential operators. In order to build a  
matrix model which has the local Lorentz invariance, we
investigate how the $so(9,1)$ Lorentz symmetry and the $u(N)$ gauge
symmetry are mixed together. We first analyze the bosonic part of
the model, and we find that the Einstein gravity is reproduced in the
classical low-energy limit. We then present a proposal to build a
matrix model which has ${\cal N}=2$ SUSY and reduces to the type 
IIB supergravity in the classical low-energy limit.

\end{abstract}
\vfill
\end{titlepage}
\vfil\eject

\section{Introduction}
 A large $N$ reduced model has been proposed as a nonperturbative
  formulation of superstring theory\cite{9612115}\cite{9705128}. It 
  is defined by the following action: 
  \begin{eqnarray}
  S = - \frac{1}{g^{2}} Tr \left( \frac{1}{4} [A_{a}, A_{b}] [A^{a},
  A^{b}] - \frac{1}{2} {\bar \psi} \Gamma^{a} [A_{a}, \psi] \right).
  \end{eqnarray}
 It is a large $N$ reduced model\cite{EguchiKawai} of 10-dimensional
 ${\cal N} =1$ super Yang-Mills theory. Here, $\psi$ is a
 10-dimensional 
 Majorana-Weyl spinor field, and $A_{a}$ and $\psi$ are $N \times N$
 hermitian matrices. 
  This model is related to the type IIB superstring theory in that
  it is at the same time obtained by the matrix regularization
  of the Green-Schwarz action of the type IIB superstring theory (IIB matrix
  model is extensively reviewed in  \cite{9908038}).   

  IIB matrix model has several evidences of describing
 gravitational interaction; such as the exchange of gravitons and
 dilatons which appears in the one-loop effective
 Lagrangian\cite{9612115}. However, it is an interesting issue to
 pursue a matrix model which describes the gravity more manifestly.
 Our goal is to build a matrix model with manifest general coordinate
 invariance, which is the fundamental principle of general relativity.
 In order to achieve this goal, we should consider a matrix model with 
 local Lorentz invariance. This issue has been
 investigated in the context of supermatrix model  in
 \cite{0102168}\cite{0103003}.   

  In these papers, the matrix model based on the
  super Lie algebra
  $u(1|16,16)$\cite{0002009}\cite{0006137}\footnote{$so(9,1)$ is a 
  subalgebra of $u(1|16,16)$. For the detailed relationship between
  the super Lie algebra and the Lorentz symmetry, see
  \cite{0003261}.} has been investigated 
  as a generalization of IIB matrix model to analyze how the 
  local Lorentz invariance is realized in a matrix model. This model
  has the following two features.
  One is the extended Lorentz symmetry, in
  which the infinitesimal parameters of $so(9,1)$ are extended to the
  elements of $u(N)$. In IIB matrix model, the information of the spacetime is
  embedded in the eigenvalues of the large $N$ matrices, and the
  general coordinate invariance is related to the $u(N)$ gauge 
  symmetry\cite{9903217}. Therefore, in order to realize the 
  coordinate-dependent Lorentz symmetry, it is natural to regard the
  parameters of $so(9,1)$ as the elements of $u(N)$. 
  The other feature of the $u(1|16,16)$ supermatrix model
  is the existence of a higher rank tensor field.
  In order to formulate a matrix model in the curved
  spacetime, a spin connection term containing  rank-3 gamma matrices
  must be included in the fermionic term. The $u(1|16,16)$ supermatrix
  model has been thus investigated as the model with higher rank tensor
  fields coupled to the fermions.

  In this paper, we inherit this idea, and attempt another
  formulation of a matrix model which manifestly possesses the local
  Lorentz invariance. The starting point is to identify the
  large $N$ matrices with differential operators\cite{tohwa01}. This idea
  stems from the interpretation of the spacetime embedded in 
  large $N$ matrices. In the twisted reduced model\cite{TEK}\cite{TEK2}, 
  the matrix $A_{a}$ represents the covariant derivative in the
  spacetime, whereas in IIB matrix model the matrix $A_{a}$ itself
  represents the coordinate of the spacetime. The relationship 
  between these two viewpoints is elucidated by expanding the matrix
  $A_{a}$ around the classical solution $p_{a}$ as
  \begin{eqnarray}
    A_{a} = p_{a} + a_{a}, 
  \end{eqnarray}
  where $[p_{a}, p_{b}] = i B_{ab}$ and $B_{ab}$ is a real
  c-number. In \cite{9908141}, it has been pointed out that IIB matrix
  model is identified with the noncommutative Yang-Mills theory by this
  expansion. The fermionic term of IIB matrix model $\frac{1}{2
  g^{2}} Tr {\bar \psi} \Gamma^{a} [A_{a}, \psi]$ reduces to the
  action of the fermion in the flat space
   \begin{eqnarray}
    \int d^{D} x {\bar \psi} (x) i \Gamma^{a} ( \partial_{a} \psi (x) +
    [a_{a} (x), \psi (x) ] ),
   \end{eqnarray}
  in the classical low-energy limit. 

  With this idea in mind, we attempt to formulate a matrix model with 
  the local Lorentz invariance, which in the classical low-energy
  limit reduces to the fermionic action on the curved spacetime  
  \begin{eqnarray}
   S_{F} = \int d^{d} x e(x) {\bar \psi}(x) i \Gamma^{a}
       {e_{a}}^{i}(x) \left(  
       \partial_{i} \psi(x) + [A_{i}(x), \psi(x)] + \frac{1}{4} \Gamma^{bc}
       \omega_{i bc} (x) \psi(x) \right). \label{AZ11ll}
  \end{eqnarray}
  Throughout this paper, the indices $a, b, c, \cdots$ and $i, j, k,
  \cdots$ both run over $0, 1, \cdots, d-1$.  The former and the
  latter denote the indices of the Minkowskian and the curved
  spacetime, respectively\footnote{In this paper, we adopt
  the following  notation of the gamma matrices: 
  \begin{eqnarray}
  \{ \Gamma^{a}, \Gamma^{b} \} = 2  \eta^{ab}, \textrm{ where }
 \eta^{ab} = \textrm{diag}(-1, +1, \cdots, +1). \nonumber 
  \end{eqnarray} 
 The gamma matrices are real and satisfy 
  \begin{eqnarray}
 (\Gamma^{a})^{\dagger} = (^{T}\Gamma^{a}) = \left\{ \begin{array}{ll}
  - \Gamma^{a} & (a=0) \\ + \Gamma^{a} & (a=1, 2, \cdots, 9)
  \end{array} \right. . \nonumber
  \end{eqnarray} 
 }. $d$ is the spacetime dimension, and we
  focus on the case in which $d=10$. In considering the
  correspondence with the matrix model, we need to absorb $e(x)$ into
  the definition of the fermionic field, since it is not $\int
  d^{d} x e(x)$ but $\int d^{d} x$ that corresponds to the trace of
  the large $N$ matrices.  And we regard the "spinor root density"  
  \begin{eqnarray}
   \Psi(x) = e^{\frac{1}{2}} (x) \psi (x) \label{AZ1rootdensity}
  \end{eqnarray}
 as the fundamental quantity. Then, the action is rewritten as
  \begin{eqnarray}
   S_{F} &=& \int d^{d} x {\bar \Psi} (x) e^{\frac{1}{2}} (x) 
    i \Gamma^{a} {e_{a}}^{i}(x) \left\{ 
      \partial_{i} (e^{-\frac{1}{2}} (x) \Psi(x) )
   + [A_{i}(x), e^{- \frac{1}{2}} (x) \Psi(x)] \right. \nonumber \\
 & & \hspace{15mm} \left. 
   + \frac{1}{4} \Gamma^{bc}
       \omega_{i bc} (x) e^{-\frac{1}{2}} (x) \Psi(x) \right\}
    \nonumber \\
    &=& \int d^{d} x \left\{ 
     {\bar \Psi} (x) i \Gamma^{a} \left[  {e_{a}}^{i} (x) \partial_{i} 
      + \frac{1}{2} {e_{c}}^{i} (x) \omega_{ica} (x)
      + {e_{a}}^{i} (x) e^{\frac{1}{2}} (x) (\partial_{i}
    e^{-\frac{1}{2}} (x) ) \right] \Psi (x)  \right. \nonumber \\
    & & \hspace{15mm} \left. + i {\bar \Psi} (x) \Gamma^{a} {e_{a}}^{i} (x)
         [A_{i}(x), \Psi (x)] 
     + \frac{i}{4} {\bar \Psi} (x) \Gamma^{a_{1} a_{2} a_{3}}
    {e_{[a_{1}}}^{i} (x) \omega_{i a_{2} a_{3}]} (x) \Psi (x)
    \right\} \nonumber \\
   &=& \int d^{d} x \left\{ {\bar \Psi} (x) i \Gamma^{a} {e_{a}}^{i} (x) 
       (\partial_{i} \Psi(x) + [A_{i} (x), {\bar \Psi} (x) ] )
     + \frac{i}{4} {\bar \Psi} (x) \Gamma^{a_{1} a_{2} a_{3}}
       {e_{[a_{1}}}^{i} (x) \omega_{i a_{2} a_{3}]} (x) \Psi
    (x) \right\}, \nonumber \\
     \label{AZ1bonafidefermion}
   \end{eqnarray}
  where we have utilized in the last equality the fact that the
  fermionic field $\Psi (x)$  is a Majorana one.  
  The corresponding matrix model is formulated as
  \begin{eqnarray}
   S'_{F} = \frac{1}{2} Tr {\bar \psi} \Gamma^{a} [A_{a}, \psi]
     +  \frac{i}{2} Tr {\bar \psi} \Gamma^{abc} \{ A_{abc}, \psi \}.
   \label{AZ11naivell}
  \end{eqnarray}
  In promoting the action (\ref{AZ1bonafidefermion}) to the matrix
  model, we have identified the covariant derivative with
  the commutator with the rank-1 matrix $A_{a}$. 
  The rank-3 term is a naive product of the spin connection and
  the fermion, and it is natural to promote the product to the
  anticommutator of the large $N$ hermitian matrices\footnote{For the
  readers' convenience, we summarize the commutation 
  relations of the hermitian and anti-hermitian
  operators. Let ${\bf H}$ and ${\bf A}$ be the set of the hermitian and
  anti-hermitian operators, respectively. When $h, h_{1}, h_{2} \in
  {\bf H}$ and $a, a_{1}, a_{2} \in {\bf A}$, their commutation
  relations are as follows: 
  \begin{eqnarray}
  [h_{1}, h_{2}] \in {\bf A}, \hspace{2mm}
  [h, a] \in {\bf H}, \hspace{2mm}
  [a_{1}, a_{2}] \in {\bf A},  \hspace{2mm}
  \{ h_{1}, h_{2} \} \in {\bf H},  \hspace{2mm}
  \{ h, a \} \in {\bf A}, \hspace{2mm}
  \{ a_{1}, a_{2} \} \in {\bf H}. \nonumber
 \end{eqnarray}
 }.
  $i$ is necessary in the rank-3 term so that the action should
  be hermitian.  The action (\ref{AZ11naivell}) is an addition of the
  rank-3 term to the fermionic term of the action of IIB matrix
  model. Especially when $\psi$ is a Majorana fermion,   
  (\ref{AZ11naivell}) is equivalent to the following action:   
   \begin{eqnarray}
   S''_{F} = Tr {\bar \psi} (\Gamma^{a} A_{a} + i \Gamma^{a_{1} a_{2}
   a_{3}} A_{a_{1} a_{2} a_{3}} ) \psi. \label{AZ11naivell2}
  \end{eqnarray}
  Let us next consider the local Lorentz transformation of this
  matrix model. Originally the local Lorentz transformation of
  the fermionic field $\delta \psi (x) = \frac{1}{4} \Gamma^{a_{1}
  a_{2}} \varepsilon_{a_{1} a_{2}} (x) \psi(x)$ should be promoted to the
  anticommutator of the hermitian matrices as $\delta \psi = \frac{1}{4}
  \Gamma^{a_{1} a_{2} } \{ \varepsilon_{a_{1} a_{2}},  \psi
  \}$. However, since it is difficult to find an action invariant
  under this transformation, we instead promote this transformation 
  to the matrix version as
  \begin{eqnarray}
   \delta \psi = \frac{1}{4} \Gamma^{a_{1} a_{2}} \varepsilon_{a_{1}
   a_{2}}  \psi, \label{AZ1alt}
  \end{eqnarray}
 and we take the action (\ref{AZ11naivell2}).
 The hermiticity of the fermionic field is now sacrificed, and the actions
 (\ref{AZ11naivell}) and (\ref{AZ11naivell2}) are no longer
 equivalent. The price we
 must pay for this alteration is that the product $A_{a} \psi$ does
 not directly correspond to the covariant derivative. 
 The local Lorentz transformation of the action (\ref{AZ11naivell2}) is  
  \begin{eqnarray}
   \delta S''_{F} = \frac{1}{4} Tr {\bar \psi} \left[ \Gamma^{a} A_{a} +
   i \Gamma^{a_{1} a_{2} a_{3}} A_{a_{1} a_{2} a_{3}}, \Gamma^{b_{1}
   b_{2}} \varepsilon_{b_{1} b_{2}} \right] \psi.
  \end{eqnarray}
  Since we are now considering the local Lorentz
  transformation and their space-time dependent parameters are promoted 
  to $u(N)$ matrices, the infinitesimal parameters  $\varepsilon_{ab}$
  are $u(N)$ matrices. Then, the commutator 
  \begin{eqnarray}
   [ i \Gamma^{a_{1} a_{2} a_{3}} A_{a_{1} a_{2} a_{3}},
  \Gamma^{b_{1} b_{2}} \varepsilon_{b_{1} b_{2}} ] 
   = \frac{i}{2} \underbrace{[ \Gamma^{a_{1} a_{2} a_{3}},
  \Gamma^{b_{1} b_{2}}]}_{\textrm{rank-3}}
       \{  A_{a_{1} a_{2} a_{3}},  \varepsilon_{b_{1} b_{2}} \}
   +  \frac{i}{2} \underbrace{\{  \Gamma^{a_{1} a_{2} a_{3}},
  \Gamma^{b_{1} b_{2}} \} }_{\textrm{rank-1, 5}}
      [  A_{a_{1} a_{2} a_{3}},  \varepsilon_{b_{1} b_{2}} ]
  \nonumber \\
  \end{eqnarray}
  must include the rank-5 gamma matrices in 10 dimensions, and the action
  (\ref{AZ11naivell2}) is not invariant 
  under the local Lorentz transformation. Repeating this procedure, we 
  find that it is necessary to introduce all the terms of odd-rank
  gamma matrices to make the action invariant.  

  The similar thing holds true of the generator of the local Lorentz
  transformation (\ref{AZ1alt}). In order for the algebra to close,
  only  the rank-2 terms are not sufficient, because the commutator  
  \begin{eqnarray}
   [\Gamma^{a_{1} a_{2}} \varepsilon_{a_{1} a_{2}}, \Gamma^{b_{1}
   b_{2}} \varepsilon'_{b_{1} b_{2}} ]
 =  \frac{1}{2} \underbrace{[ \Gamma^{a_{1} a_{2}}, \Gamma^{b_{1}
   b_{2}} ]}_{\textrm{rank-2}}
     \{  \varepsilon_{a_{1} a_{2}},  \varepsilon'_{b_{1} b_{2}} \}
  + \frac{1}{2} \underbrace{\{ \Gamma^{a_{1} a_{2}}, \Gamma^{b_{1}
   b_{2}} \} }_{\textrm{rank-0, 4}}
     [  \varepsilon_{a_{1} a_{2}},  \varepsilon'_{b_{1} b_{2}} ]
  \end{eqnarray}
  includes the rank-4 gamma matrices. Likewise, we
  find that the algebra of the local Lorentz transformation must
  include all the even-rank gamma matrices.

  Our formulation of the matrix model is based on this observation,
  and we analyze a matrix model which includes all odd
  ranks of 10-dimensional gamma matrices coupled to the fermion from
  the beginning. We identify these matrices with differential
  operators, and investigate whether our matrix model includes the
  supergravity in the classical low-energy limit. 

  We mention the relation between our new proposal and the original IIB 
  matrix model. In the quantum field theory, some different
  models which share the same symmetry are equivalent in the continuum
  limit, known as universality.
  We expect that the similar mechanism may hold true of 
  large $N$ matrix models and hence that various matrix models may
  have the same large $N$ limit. Thus, there is a possibility that our 
  new model is equivalent to IIB matrix model.
  We can expect that both IIB matrix model and our new model are
  equally authentic constructive definition of superstring theory in
  the large $N$ limit. 
  Even if IIB matrix model is not an eligible framework, we expect
  that our model may be closer to the true theory than IIB matrix
  model, because our model has larger symmetry.
 
  This paper is organized as follows. In Section 
  2, we analyze the bosonic term of the matrix model as the
  preliminary step, and we clarify that the bosonic part reduces to
  the Einstein gravity in the classical low-energy limit.
  In Section 3, we consider the model coupled to a fermionic field. We
  present a proposal to formulate a matrix model which has
  ${\cal N}=2$ SUSY algebra and reduces to the type IIB
  supergravity. Section 4 is devoted to the discussion and the outlook. 

  \section{Bosonic part of the matrix model}
  Before going to the analysis of the supersymmetric matrix model, we
  first investigate the bosonic part:
   \begin{eqnarray}
     S = Tr_{N \times N} \left[ 
   tr_{32 \times 32} V(m^{2})  \right]. \label{AZ2action}
  \end{eqnarray}
 The uppercase $Tr$ and the lowercase  $tr$ denote the trace for the
 $N \times N$ and $32 \times 32$ matrices, respectively.
 The large $N$ matrix $m$ consists of all odd-rank gamma
 matrices in 10 dimensions:  
 \begin{eqnarray}
   m = m_{a} \Gamma^{a} 
     + \frac{i}{3!} m_{a_{1} a_{2} a_{3}} \Gamma^{a_{1} a_{2} a_{3}}
     - \frac{1}{5!} m_{a_{1} \cdots a_{5}} \Gamma^{a_{1} \cdots a_{5}}
     - \frac{i}{7!} m_{a_{1} \cdots a_{7}} \Gamma^{a_{1} \cdots a_{7}}
     + \frac{1}{9!} m_{a_{1} \cdots a_{9}} \Gamma^{a_{1} \cdots
     a_{9}}, \nonumber \\ \label{AZ2mex}
 \end{eqnarray}
 where  $m_{a_{1} \cdots a_{2n-1}}$ are large $N$ hermitian matrices.
 $m$ satisfies 
  \begin{eqnarray}
   \Gamma^{0} m^{\dagger} \Gamma^{0} = m,
  \end{eqnarray}
 and thus the action (\ref{AZ2action}) is hermitian.
 Odd powers of $m$ must not appear in the action, since they would
 transform the fermion to that of the opposite chirality. 
 The function $V(m^{2})$ is discussed in Section \ref{AZpotential}. 

 \subsection{Identification of large $N$ matrices with differential
 operators} 
 We identify the space of large $N$ matrices with that of the
 differential operators\cite{tohwa01}. 
 By this identification, we can describe the differential
 operators on an arbitrary spin bundle over an arbitrary manifold in the
 continuum limit simultaneously, because they are embedded in the
 space of large $N$ matrices (some simple examples are provided in
 Appendix \ref{diffmatrix}). In the following, from the big space of
 large $N$ matrices, we pick up a subspace consisting of the
 differential operators over one manifold.
 We  regard $m$ as the differential operators
 over this manifold. Then, we analyze the effective theory of the
 fields appearing in the expansion of the differential 
 operators (the explicit form is given later in
 (\ref{AZ2bosonexpansion})). 

 The space of the differential operator is infinite-dimensional,
 and in general the trace $Tr$ for this space is divergent. But, 
 as we elucidate later, we choose a function $V(m^{2})$ which
 decreases exponentially, and thus the trace is finite.  
 We want to identify the dimensionless matrix $m$ with something
 similar to the Dirac
 operator, and we need to introduce $D$ as an extension of the Dirac
 operator in the curved space acting on the "spinor root density"
 (\ref{AZ1rootdensity}). $D$ has the dimension
 $[\textrm{(length)}^{-1}]$, and $m$ is expressed as  
 \begin{eqnarray}
   m &=& \tau^{\frac{1}{2}} D, \textrm{ where }
   \label{AZ2verydef} \\
    D &=& A_{a} \Gamma^{a} 
   + \frac{i}{3!} A_{a_{1} a_{2} a_{3}} \Gamma^{a_{1} a_{2} a_{3}} 
   - \frac{1}{5!} A_{a_{1} \cdots a_{5}} \Gamma^{a_{1} \cdots a_{5}}
   - \frac{i}{7!} A_{a_{1} \cdots a_{7}} \Gamma^{a_{1} \cdots a_{7}} 
   + \frac{1}{9!} A_{a_{1} \cdots a_{9}} \Gamma^{a_{1} \cdots a_{9}},
   \nonumber \\
  \label{AZ2expansionD}
  \end{eqnarray}
 where $\tau$ has the dimension $[\textrm{(length)}^{2}]$.
 $\tau$ is similar to the Regge slope $\alpha'$ in string theory,
 and is introduced as a reference scale.
 This parameter $\tau$ is not a cut-off parameter.
 $V(m^{2})$ is an exponentially decreasing function, and
 the damping factor is supplied by the action itself.
 Since $V(m^{2})$ is a function of the dimensionless quantity $m$,
 $\tau$ represents a damping scale.
 When we approximate the differential operators by finite $N$
 matrices, an $N$-dependent ultraviolet cut-off naturally appears.
 When we take $N$ to infinity, this cut-off becomes
 infinitely small. But the scale $\tau$ is completely
 independent of this ultraviolet cut-off, and takes a constant value
 even in the large $N$ limit.  Thus, we can fairly take $N$ to
 infinity and identify the large $N$ matrices with the differential
 operators, at least when we investigate the effective theory at tree
 level. In this sense, our model {\it differs} from the induced gravity.    
 
  $A_{a_{1} \cdots a_{2n-1}}$ $(n=1, 2, \cdots, 5)$ are hermitian
 differential operators and are expanded by the number of the
  derivatives. Because of the 
 hermiticity, they are expanded as the anticommutator of
 differential operators and real functions:
  \begin{eqnarray}
  A_{a_{1} \cdots a_{2n-1}} = a_{a_{1} \cdots a_{2n-1}} (x) 
    + \sum_{k=1}^{\infty} \frac{i^{k}}{2} \{ \partial_{i_{1}} \cdots
    \partial_{i_{k}} , {a^{(i_{1} \cdots i_{k})}}_{a_{1} \cdots
    a_{2n-1}} (x) \}. \label{AZ2bosonexpansion}
  \end{eqnarray}
  The indices in the parentheses are symmetric, while the other
 indices are antisymmetric. 
 Since $D$ is an extension of the Dirac operator in the curved space,
 we find it natural to identify the function ${a_{a}}^{(i)} (x)$ with
 the vielbein of the background metric. Now, the operator $D$ is
 expanded as  
   \begin{eqnarray}
   D &=& e^{\frac{1}{2}} (x) \left[ i {e_{a}}^{i} (x) \Gamma^{a} 
   \left( \partial_{i} + \frac{1}{4} \Gamma^{bc} \omega_{ibc} (x)
   \right) \right] e^{- \frac{1}{2}} (x) \nonumber \\
     &+& (\textrm{higher-rank terms}) + (\textrm{higher-derivative
   terms}).  \label{AZ2ferm}
  \end{eqnarray}
 The other coefficients ${a^{(i_{1} \cdots i_{k})}}_{a_{1} \cdots
  a_{2n-1}} (x)$  are regarded as the matter fields, and a simple
  dimensional analysis reveals that they have the dimension   
  \begin{eqnarray}
   {a^{(i_{1} \cdots i_{k})}}_{a_{1} \cdots a_{2n-1}} (x) \sim
   [\textrm{(length)}^{-1+k} ]. \label{AZ2dimension}
  \end{eqnarray}

 \subsection{Local Lorentz invariance} 
 We next investigate the local Lorentz symmetry of this action. The local
 Lorentz transformation is 
  \begin{eqnarray}
   \delta m &=& [ m, \varepsilon], \textrm{ where } \nonumber \\
   \varepsilon &=&  - i \varepsilon_{\emptyset} 
  + \frac{1}{2!} \Gamma^{a_{1} a_{2}} \varepsilon_{a_{1} a_{2}} 
  + \frac{i}{4!} \Gamma^{a_{1} \cdots a_{4}} \varepsilon_{a_{1} \cdots
   a_{4}}  
  - \frac{1}{6!} \Gamma^{a_{1} \cdots a_{6}} \varepsilon_{a_{1} \cdots
   a_{6}} \nonumber \\
  & & - \frac{i}{8!} \Gamma^{a_{1} \cdots a_{8}} \varepsilon_{a_{1} \cdots
    a_{8}}
  + \frac{1}{10!} \Gamma^{a_{1} \cdots a_{10}} \varepsilon_{a_{1} \cdots
    a_{10}}. \label{AZ2locallorentz}
  \end{eqnarray}
 The large $N$ matrix $\varepsilon$ satisfies 
  \begin{eqnarray}
    \Gamma^{0} \varepsilon^{\dagger} \Gamma^{0} = \varepsilon,
  \end{eqnarray}
 and represents the local Lorentz transformation, with
 $\varepsilon_{a_{1} \cdots a_{2n}}$ $(n=0, \cdots, 5)$ being
 hermitian. As we have observed in Section 1,  
 all the even-rank gamma matrices are necessary in order for the
 algebra to close. The invariance of the action is verified as 
  \begin{eqnarray}
    \delta S  = Tr \left[ tr \left( 
  2  V'(m^{2}) m \left[ m, \varepsilon \right] \right)
   \right]  = 0, 
  \end{eqnarray}
 where $V'(x)$ denotes  $V'(x) = \frac{\partial V(x)}{\partial x}$. 
 The cyclic property of the trace still holds true of the space of the 
 differential operators, if we assume that all the coefficients damp
 rapidly at infinity; i.e. 
  \begin{eqnarray}
 \lim_{|x| \to + \infty} {a^{(i_{1} \cdots i_{k})}}_{a_{1} \cdots
 a_{2n-1}} (x) = 0. \label{AZ2assumption} 
  \end{eqnarray}
  Under this assumption, the trace of the commutator between the
  function and the differential operator is 
  \begin{eqnarray}
  & &   Tr ( [ \partial_{j}, {a^{(i_{1} \cdots i_{k})}}_{a_{1} \cdots
 a_{2n-1}}  (x) ] ) \nonumber \\
 &=&  \int d^{d} x \langle x | (\partial_{j} {a^{(i_{1} \cdots
 i_{k})}}_{a_{1} \cdots a_{2n-1}} (x) ) | x \rangle  
 = \int d^{d} x (\partial_{j} {a^{(i_{1} \cdots i_{k})}}_{a_{1} \cdots
 a_{2n-1}} (x) ) \langle x | x \rangle.
  \end{eqnarray}
  The surface term of this integral vanishes due to the assumption
  (\ref{AZ2assumption}), and the cyclic property of the trace is
  ensured even for the trace of differential operators.

 \subsection{Determination of $V(m^{2})$}
 \label{AZpotential}
 We next discuss the condition $V(m^{2})$ should
 satisfy. In this section, we set the background metric to be flat for 
 brevity.  We require that the differential operator in the flat space
 $m_{0} = i \tau^{\frac{1}{2}} \Gamma^{a} \partial_{a}$ should be a
 classical solution of this matrix model. To investigate this
 condition, we consider the Laplace transformation of the function
 $V(u)$     
  \begin{eqnarray}
   V(u) = \int^{\infty}_{0} ds g(s) e^{- su}. \label{AZ2laplace}
  \end{eqnarray}
 The Laplace transformation (\ref{AZ2laplace}) enables us to relate
 the trace $Tr [ tr V(m^{2}) ]$ with the Seeley de Witt expansion of
 the trace of $e^{-\tau D^{2}}$ \cite{9606001} \cite{publishorperish},
 which is the expansion around $m_{0}^{2} = - \tau \partial_{a}
 \partial^{a}$. The bosonic part of the action is thus expressed as 
 \begin{eqnarray}
  Tr [ trV(m^{2})] &=& \int^{\infty}_{0} ds g(s) Tr[ tr e^{-s \tau
 D^{2}}] \nonumber \\
 &=& \int^{\infty}_{0} ds g(s) \left( \int \frac{d^{d} x}{(2
 \pi)^{\frac{d}{2} } } 
  [ \cdots + (s \tau)^{- \frac{d}{2} - 2} {\cal A}_{-2} (x) 
   + (s \tau)^{- \frac{d}{2} - 1} {\cal A}_{-1}
 (x) \right. \nonumber \\
  & & \left. \hspace{10mm} 
   + (s \tau)^{- \frac{d}{2}} {\cal A}_{0} (x)
   + (s \tau)^{- \frac{d}{2} + 1} {\cal A}_{1} (x)
   + (s \tau)^{- \frac{d}{2} + 2} {\cal A}_{2} (x) + \cdots ] \right)
 \nonumber \\
  &=& \int \frac{d^{d} x}{(2 \pi \tau)^{\frac{d}{2}}} 
  \left(  \sum_{k=-\infty}^{\infty} \left(\int^{\infty}_{0} ds g(s)
 s^{-\frac{d}{2} + k} \right) \tau^{k} {\cal
 A}_{k} (x) \right),  \label{AZ2trace}
 \end{eqnarray}
 where $k$ runs over all the integers\footnote{A dimensional analysis
 shows that the terms of the order ${\cal O}(\tau^{- \frac{d}{2} +
 l})$ for half integer $l$ do not appear in this expansion. From
 (\ref{AZ2dimension}), the  dimension of the term
 $\prod_{i=1}^{n} (\partial_{j_{1}} \cdots 
 \partial_{j_{p_{i}} } {a^{(k_{1} \cdots k_{l_{i}})}}_{b_{1} \cdots
 b_{2m_{i} - 1}} (x) )$ is 
 $[(\textrm{length})^{{\cal D}}]$, where ${\cal D} =  - n +
 \sum_{i=1}^{n} ( - p_{i} + l_{i} ) $. 
 This term is included in the Seeley de Witt coefficient 
 ${\cal A}_{\frac{{\cal D}}{2}} (x)$.  On the other hand, the number
 of the indices in this term is 
 $\sum_{i=1}^{n} (2 m_{i} - 1 + p_{i} + l_{i} )$, which is equal to
 ${\cal D}$ modulo 2. When ${\cal D}$ is odd, the total
 number of the indices is also odd and this term cannot contract the
 indices to constitute a scalar. Therefore, the coefficient ${\cal
 A}_{l} (x)$ for half integer $l$ is prohibited due to the Lorentz
 invariance of the action. However, they may survive when $d$ is odd, with
 the antisymmetric tensor $\epsilon_{i_{1} \cdots  i_{d}}$ being
 accompanied.}. $A_{k} (x)$ are the 
 coefficients of the Seeley de Witt expansion of the trace $Tr
 [ tr(e^{- s \tau D^{2}})]$. And there appear the
 coefficients  $A_{k} (x)$ for negative $k$, too, because 
 the fields ${a^{(i_{1} \cdots i_{k})}}_{a_{1} \cdots a_{2n-1}} (x)$
 possess the dimension (\ref{AZ2dimension}).

 In order for $m_{0}$ to be a classical solution, the linear term of
 the fluctuation around $m_{0}$ should vanish in the action. 
 Since only the scalar fields can constitute a Lorentz invariant linear
 term, we focus on the linear terms with the indices contracted as
 ${a_{a}}^{(a i_{1} i_{1} \cdots i_{l} i_{l})} (x)$ ($l=0, 1, 2,
 \cdots$). 
 Since ${a_{a}}^{(i)} (x)$ is identified with the vielbein, the
 cancellation of the linear term ${a_{a}}^{(a)} (x)$ means the
 absence of the cosmological constant in this matrix model.
 The linear terms of their derivatives 
 $(\partial_{j_{1}} \cdots \partial_{j_{m}} {a_{a}}^{(a i_{1} i_{1} \cdots
 i_{l} i_{l})} (x) )$ disappear after integrating in the action.
 A simple dimensional analysis indicates that the linear 
 terms of the fields ${a_{a}}^{(a i_{1} i_{1} \cdots
 i_{l} i_{l})} (x)$ are included in ${\cal A}_{-l} (x)$. 
 Therefore, we demand that the Seeley de Witt coefficients ${\cal
 A}_{-l} (x) $ $(l=0, 1, 2, \cdots)$ should be canceled in the
 action. 

 The condition for $m_{0}$ to be a classical solution is thus
 translated into  the following statement:
  \begin{eqnarray}
   \int^{\infty}_{0} ds g(s) s^{- \frac{d}{2} - n} = 0, \textrm{ for }
   n=0, -1, -2, \cdots. \label{AZ2condition1}
  \end{eqnarray}
 Now, let us rewrite the condition (\ref{AZ2condition1}) in terms of
 the function $V(u)$. Noting 
  \begin{eqnarray}
   \int^{\infty}_{0} du V(u) u^{\alpha - 1} 
 = \int^{\infty}_{0} du \int^{\infty}_{0} ds g(s) e^{-su} u^{\alpha
 -1}
 = \Gamma (\alpha) \int^{\infty}_{0} ds g(s) s^{-\alpha},
  \end{eqnarray}
 we discern that the function $V(u)$ must satisfy\footnote{Since
 $\frac{d}{2} + n > 0$ for $n= -1, 0, 1, 2, \cdots$, we
 need not worry about the possibility that $\Gamma (\frac{d}{2} + n +
 1)$ vanishes.}
  \begin{eqnarray}
   \int^{\infty}_{0} du u^{\frac{d}{2} + n} V(u) = 0, \textrm{ for } 
   n = -1, 0, 1, 2, \cdots. \label{AZ22acidtest}
  \end{eqnarray}
  
 There are various choices for such functions, and one example is the
 following function:\footnote{We can verify that
 this function satisfies (\ref{AZ22acidtest}) as
 follows\cite{Hausdorff}. We note that 
  $\int^{\infty}_{0} dy y^{m} e^{- ay} = m! a^{-m-1} $ for $a = \exp
 (\frac{i \pi}{4}) = \frac{1+i}{\sqrt{2}}$ and $m = 0, 1, 2,
 \cdots$. This is a real number  when $m-3$ is a multiple of 4. 
 Taking the imaginary part of the both hand sides, we obtain  
 $\int^{\infty}_{0} dy y^{4n+3} \sin ( \frac{y}{\sqrt{2}} ) \exp ( - 
 \frac{y}{\sqrt{2}} ) = 0$, for $n=0, 1, 2, \cdots$. We make a
 substitution $u = \frac{y^{4}}{4}$ and integrate by part to verify
 that (\ref{AZ22example1}) satisfies
 (\ref{AZ22acidtest}). Another choice is the function 
  $V(u) = \frac{\partial^{\frac{d}{2} -1} (\sin \sqrt{u})}{\partial
 u^{\frac{d}{2}-1}} $, which corresponds to the limit $a = \exp 
 ( \frac{i \pi}{2} )$. In this case, we need to introduce a proper
 damping factor.}
  \begin{eqnarray}
    V_{0}(u) = \frac{\partial^{\frac{d}{2} -1} (\exp(- u^{\frac{1}{4}}
    ) \sin(u^{\frac{1}{4}} ) ) }{\partial
    u^{\frac{d}{2} - 1} }.  \label{AZ22example1} 
  \end{eqnarray}

 Going back to the general case, when  $V(m^{2})$ is chosen in order
 that $m_{0} = i \tau^{\frac{1}{2}} \Gamma^{a} \partial_{a}$ becomes a
 classical solution, we understand that the action reduces to the
 Einstein gravity in the classical low-energy limit. 
 The absence of the linear term ${a_{(a)}}^{a} (x)$ ensures that the
 graviton is massless, since the mass terms (including the
 cross-terms with the other fields) vanish due to the general
 coordinate invariance. Then, the Einstein gravity is derived from
 the Seeley de Witt expansion of the trace $Tr[ tr( e^{- \tau
 D^{2}})]$ when we bequeath only  the curved space Dirac operator 
 in the expansion (\ref{AZ2ferm}):\footnote{The square density
 $e^{\frac{1}{2}} (x)$ can be treated by using the cyclic rule of the
 trace.}   
  \begin{eqnarray}
   Tr[ tr( e^{- \tau D^{2}})] = \int d^{d} x \frac{32}{(2 \pi
   \tau)^{\frac{d}{2}}} e(x) \left( \tau \frac{R(x)}{6} + \cdots \right),
  \end{eqnarray}
 where $32$ comes from  the trace $tr {\bf 1}_{32 \times 32}$. 

It is obvious from the dimensional analysis that this
 term is included in ${\cal A}_{1} (x)$. When ${\cal A}_{1} (x)$
 survives in  the action, i.e. $\int^{\infty}_{0} ds g(s) s^{-
 \frac{d}{2} + 1} \neq 0$ or equivalently $\int^{\infty}_{0} du V(u)
 u^{\frac{d}{2} - 2} \neq 0$, the trace $Tr [ tr V(m^{2})]$ includes the
 Einstein gravity in the Seeley de Witt expansion.  

 We next investigate the mass terms and the kinetic terms of the
 matter fields. A dimensional analysis shows that the mass terms and
 the kinetic terms (including the cross terms) are included in the
 following Seeley de Witt coefficients:  
  \begin{eqnarray}
   & & \textrm{mass terms: } {a^{(i_{1} \cdots i_{k})}}_{a_{1} \cdots 
   a_{2n-1}} (x) {a^{(j_{1} \cdots j_{l})}}_{a_{1} \cdots a_{2n-1}}
   (x) \in {\cal A}_{1 - \frac{k+l}{2}} (x),  \\
   & &  \textrm{kinetic terms: } (\partial_{k_{1}} {a^{(i_{1} \cdots
   i_{k})}}_{a_{1} \cdots  a_{2n-1}} (x) )
   (\partial_{k_{2}} {a^{(j_{1} \cdots j_{l})}}_{a_{1} \cdots a_{2n-1}}
   (x) ) \in {\cal A}_{2 - \frac{k+l}{2}} (x),
  \end{eqnarray}
 where $k+l$ is an even number.
 In particular, the mass terms and the kinetic terms of the odd-rank
 anti-symmetric fields $a_{a_{1} \cdots  a_{2n-1}} (x)$ are included in 
 ${\cal A}_{1} (x)$ and ${\cal A}_{2} (x)$,
 respectively. Generically\footnote{The computation of the trace $Tr
 [ tr(e^{- \tau D^{2}})]$  indicates that the  mass term 
 $a^{a} (x) a_{a} (x)$ vanishes, but $a_{a} (x)$ is presumably massive 
 because there is no reason that the cross-terms
 $a^{a} (x) {a^{(j_{1} \cdots j_{2l})}}_{a} (x)$ also vanish.}, 
 these fields are massive and are decoupled in the classical
 low-energy limit. 

 Next, we consider the fields ${a^{(i)}}_{a_{1} \cdots a_{2n-1}}
 (x)$. As we will explain in Section 3, especially the fields
 ${a^{(i)}}_{i a_{1} \cdots a_{2n}} (x)$ (with $n=1, 2, 3, 4$) are
 identified with the anti-symmetric tensor fields in the type IIB
 supergravity. Their mass terms are included in ${\cal A}_{0} (x)$, 
 whereas the kinetic terms reside in ${\cal A}_{1} (x)$. These 
 fields are therefore massless. 
 This observation provides a positive evidence in constructing a
 matrix model which reduces to the type IIB supergravity, which is the 
 goal of the discussion of the next section. 

 However, it is not clear whether the higher-spin fields
 ${a^{(i_{1} \cdots  i_{k})}}_{a_{1} \cdots a_{2n-1}} (x)$
 ($k=2, 3, \cdots$) are massive, since the mass terms and the
 kinetic terms belong to the coefficients  $A_{1-k} (x)$ and $A_{2-k}
 (x)$ respectively and both of them vanish in the action. 
  
 \section{Supersymmetric Action}
  In the previous section, we have investigated the bosonic part of
  the matrix model, and we have elucidated that the Einstein gravity is
  derived from the bosonic action in the classical low-energy limit.

    We next add the fermionic term and present a proposal to construct
    a matrix model which has ${\cal N}=2$ SUSY and 
    reduces to the type IIB  supergravity in the classical low-energy
    limit. If we succeed in building 
    such a matrix model, this will be an extension of IIB matrix model 
    with the gravity encoded more manifestly and provide a
    strong evidence that a matrix model is an eligible framework to
    describe the gravitational interaction. The suggestion in this 
    section is rather conjectural. We start with the following
    action: 
  \begin{eqnarray}
  S_{S} &=& Tr[ tr V_{S} (m^{2}) ] + Tr {\bar \psi} m
  \psi. \label{AZ3action} 
  \end{eqnarray}
 Here, $m$ is a large $N$ matrix already defined in (\ref{AZ2mex}).
 $m$ is identified with the differential operators and  
 expanded by the number of the derivatives in the same
 way as is  explained in the previous section. 

 $\psi$ is a Weyl fermion, but is no longer hermitian.
 As we have observed in Section 1, it is difficult to construct
 a matrix model which is invariant under the local Lorentz
 transformation of the type $\delta \psi = \frac{1}{4} \Gamma^{ab} 
 \{ \varepsilon_{ab},  \psi \}$. Then, we define the local Lorentz
 transformation of the matrix model as $\delta \psi = \varepsilon
 \psi$ and sacrifice the hermiticity of the fermionic field. This
 fermionic field is also identified with differential operators and
 expanded as    
  \begin{eqnarray}
   \psi = \left( \chi(x) + \sum_{l=1}^{\infty}  
  \chi^{(i_{1} \cdots i_{l})} (x) \partial_{i_{1}} \cdots
  \partial_{i_{l}} \right) e^{- (\tau D^{2})^{\alpha}},
 \label{AZ3expansionf} 
  \end{eqnarray}
 where the fermionic fields $\chi^{(i_{1} \cdots i_{l})} (x)$ possess
 the dimension $[\textrm{(length)}^{l}]$.
 We temporarily introduce the damping factor $e^{- (\tau
 D^{2})^{\alpha}}$ 
 so that the trace $Tr {\bar \psi} m \psi$ should be finite. The
 power $\alpha$ is chosen in accordance with the function
 $V_{S}(m^{2})$ and, for example, when we choose the function
 (\ref{AZ22example1}), the power is $\alpha = \frac{1}{4}$.  

 This action is invariant under the local Lorentz transformation
  \begin{eqnarray}
   \delta m = [m, \varepsilon], \hspace{3mm} 
   \delta \psi = \varepsilon \psi,
  \end{eqnarray}
 where $\varepsilon$ is already defined in 
 (\ref{AZ2locallorentz}). The transformation of ${\bar \psi}$ is
 readily seen to be $\delta {\bar \psi} = - 
 {\bar \psi} \varepsilon$. The local Lorentz invariance is  
 verified as
 \begin{eqnarray}
 \delta S_{S} = 2 Tr[ tr( V_{S}'(m^{2}) m  [m, \varepsilon])]  
  + Tr [ tr( {\bar \psi}[m, \varepsilon] \psi)] = 0,
  \label{AZ3locallorentz} 
 \end{eqnarray}
 when we assume that the coefficients ${a^{(i_{1} \cdots i_{k})}}_{a_{1}
 \cdots a_{2n-1}}(x)$ and $\chi^{(i_{1} \cdots i_{k})} (x)$ damp
 rapidly at infinity. 

 The function $V_{S} (m^{2})$ is to be determined so that it satisfies 
 the following conditions. First, the differential operator in the flat 
 space $m_{0} = i \tau^{\frac{1}{2}} \Gamma^{a} \partial_{a}$ must be
 a classical solution, and $V_{S} (m^{2})$ must satisfy the 
 criterion (\ref{AZ22acidtest}).
 Second, in order for the model to reduce to the type IIB supergravity
 in the classical low-energy limit, only the fields ${a^{(i)}}_{i
 a_{1} a_{2} \cdots a_{2n}} (x)$ $(n=1, 2, 3, 4)$, which are
 interpreted as the anti-symmetric even-rank fields in
 the type IIB supergravity, must be massless and the other fields must be
 massive.  It is not clear whether (\ref{AZ22example1}) is an eligible
 function, because we have yet to understand whether the higher-spin
 fields are massive. This problem will be reported in more detail
 elsewhere. 

 We next consider the structure of the supersymmetry of this matrix
 model. The SUSY transformation is given by 
  \begin{eqnarray}
   \delta_{\epsilon} m = \epsilon {\bar \psi} + \psi {\bar \epsilon}, 
   \hspace{3mm}
   \delta_{\epsilon} \psi = 2 V_{S}'(m^{2}) \epsilon. 
  \end{eqnarray}
  It readily follows that the SUSY transformation of ${\bar \psi}$ is 
  \begin{eqnarray}
   \delta_{\epsilon} {\bar \psi} = 2 {\bar \epsilon} V_{S}'(m^{2}).
  \end{eqnarray}
 The invariance of the action under this SUSY transformation is
 verified as
  \begin{eqnarray}
  \delta_{\epsilon} S_{S} &=&  Tr \left[ tr \left( 
   \left( 2 V_{S}'(m^{2}) m  ( \epsilon {\bar \psi} + \psi {\bar
   \epsilon})  \right)
    + {\bar \psi} ( \epsilon {\bar \psi} + \psi {\bar \epsilon}) \psi
  \right. \right. \nonumber \\
   & & \hspace{5mm} \left. \left.
  +  2 {\bar \psi} m V_{S}'(m^{2}) \epsilon
  +  2 {\bar \epsilon}  m V_{S}'(m^{2}) \psi \right) \right] = 0.
 \label{AZ3susyinvariance}
  \end{eqnarray} 
 We analyze the commutator of this SUSY transformation especially when
 it is possible to perform the Taylor expansion of $V_{S}(u)$ around $u=0$ 
 as
 \begin{eqnarray}
  V_{S} (u) = \sum_{k=1}^{\infty} \frac{a_{2k}}{2k} u^{k}, \label{AZ3taylor}
 \end{eqnarray}
 in order to simplify the analysis using the equations of motion
  \begin{eqnarray}
   \frac{\partial S_{S}}{\partial {\bar \psi}} = 2 m \psi = 0,
   \hspace{3mm} 
   \frac{\partial S_{S}}{\partial \psi} = 2 {\bar \psi} m = 0.
  \label{AZ3EOM}
  \end{eqnarray}
 The following analysis is not applied to the function
 (\ref{AZ22example1}), because the Taylor expansion
 around $u=0$ is impossible\footnote{The invariance of the action
 under the local Lorentz transformation (\ref{AZ3locallorentz}) and
 the SUSY transformation (\ref{AZ3susyinvariance}) holds true without
 assuming the Taylor expansion around $u=0$.}. 
 However, in the case of (\ref{AZ3taylor}), the commutator can be
 simplified as  
  \begin{eqnarray}
  & & [\delta_{\epsilon}, \delta_{\xi}] m =
   2  [\xi {\bar \epsilon} - \epsilon {\bar \xi}, V_{S}'(m^{2} )],
     \label{AZ3comm-m} \\
  & & [\delta_{\epsilon}, \delta_{\xi}] \psi =
  2 \psi \left( {\bar \epsilon} m \frac{V_{S}'(m^{2}) - V_{S}'(0)}{m^{2}}
  \xi - {\bar \xi} m \frac{V_{S}'(m^{2}) - V_{S}'(0)}{m^{2}} \epsilon
  \right), 
  \label{AZ3comm-psi} 
  \end{eqnarray}
 The explicit derivation is presented in Appendix. \ref{AZexplicit3}.

 In order to see the structure of the ${\cal N} = 2$ SUSY, we separate 
 the SUSY parameters into the hermitian and the antihermitian parts as
  \begin{eqnarray}
   \epsilon = \epsilon_{1} + i \epsilon_{2}, \hspace{3mm}
   \xi = \xi_{1} + i \xi_{2},
  \end{eqnarray}
 where $\epsilon_{1}$, $\epsilon_{2}$, $\xi_{1}$ and $\xi_{2}$ are
 Majorana-Weyl fermions. For simplicity, we assume that
 $\epsilon$ and $\xi$ are c-numbers and that the background metric is 
 flat. We now clarify that the translation of $A_{a}$ comes from 
 the quartic term of $m$ in the Taylor expansion (\ref{AZ3taylor}):
  \begin{eqnarray}
  & & [ \delta_{\epsilon}, \delta_{\xi}] A_{a}
   = \frac{1}{16} tr ( [\delta_{\epsilon}, \delta_{\xi}] m \Gamma^{a}
   ) \nonumber \\
  & & \hspace{15mm} = 
   \frac{1}{16} \sum_{k=2}^{\infty} a_{2k} tr ( \xi {\bar \epsilon} m^{2k-2}
   \Gamma^{a} - \epsilon {\bar \xi} m^{2k-2} \Gamma^{a}
   - m^{2k-2} \xi {\bar \epsilon} \Gamma^{a}
   + m^{2k-2} \epsilon {\bar \xi} \Gamma^{a} ) \nonumber \\
  & & \hspace{15mm} = 
   \frac{1}{16} \sum_{k=2}^{\infty} a_{2k} ( {\bar \xi} [m^{2k-2},
   \Gamma^{a}] \epsilon - {\bar \epsilon} [m^{2k-2}, \Gamma^{a}] \xi ) 
   \nonumber \\
 & & \hspace{15mm} = 
   \frac{a_{4}}{16}  ( {\bar \xi} [\Gamma^{b_{1}} \Gamma^{b_{2}},
   \Gamma^{a}] \epsilon - {\bar \epsilon} [\Gamma^{b_{1}} \Gamma^{b_{2}},
   \Gamma^{a}] \xi ) A_{b_{1}} A_{b_{2}} + \cdots 
   = \frac{a_{4}}{16} ( {\bar \xi} \Gamma^{i} \epsilon
    - {\bar \epsilon} \Gamma^{i} \xi ) [A_{i}, A_{a}] + \cdots
 \nonumber \\
 & & \hspace{15mm} = 
   \frac{a_{4}}{8} ( {\bar \xi}_{1} \Gamma^{i} \epsilon_{1}
                    +  {\bar \xi}_{2} \Gamma^{i} \epsilon_{2} )
      [A_{i}, A_{a} ] + \cdots, \label{AZ3trans-a}
\end{eqnarray}
 where we have utilized the fact that the SUSY parameters $\epsilon$
 and $\xi$ are c-numbers and thus that the matrix $m$ commutes with 
 the SUSY parameters. Let us  concentrate on the SUSY
 transformation of the field $a_{a} (x)$. The commutator  $[A_{i}, A_{a}]$
 represents the translation of $a_{a} (x)$ and its gauge
 transformation:
  \begin{eqnarray}
  [A_{i}, A_{a}] &=& [i \partial_{i} + a_{i} (x), i \partial_{a} + a_{a} 
  (x) ] + \cdots \nonumber \\
  &=& \underbrace{i (\partial_{i} a_{a} (x) )}_{\textrm{translation}} 
  \underbrace{ - i (\partial_{a} a_{i} (x) )
    + [a_{i} (x), a_{a} (x) ]}_{\textrm{gauge transformation}} + \cdots.
  \end{eqnarray}
 Therefore, we discern that the commutator of the SUSY transformations
 constitutes a bona fide translation for the vector fields.

 However, this is not the case with the fermionic fields.
 The commutator is computed as 
\begin{eqnarray}
 & &    [\delta_{\epsilon}, \delta_{\xi}] \psi
  =  -  \sum_{k=2}^{n} a_{2k} \psi ( {\bar \xi} m^{2k-3} \epsilon
      - {\bar \epsilon} m^{2k-3} \xi ) + \cdots
  = -  a_{4} ( {\bar \xi} \Gamma^{j} \epsilon - {\bar
   \epsilon} \Gamma^{j} \xi ) \psi A_{j} + \cdots \nonumber \\
 & &  \hspace{15mm} = - 2 a_{4} ({\bar \xi}_{1} \Gamma^{j} \epsilon_{1}
      + {\bar \xi}_{2} \Gamma^{j} \epsilon_{2} ) \psi A_{j} + \cdots.
  \label{AZ3trans-psi}
  \end{eqnarray}
 We explore the term $\psi A_{i}$  more carefully: 
  \begin{eqnarray}
  \psi A_{j} &=& i \psi \partial_{j} + \cdots 
    = \left( \chi (x) \partial_{j} + \sum_{l=1}^{\infty} \chi^{(i_{1}
    \cdots i_{l})} (x) \partial_{i_{1}} \cdots \partial_{i_{l}}
    \partial_{j} \right) e^{- (\tau D^{2})^{\alpha}} + \cdots, 
  \end{eqnarray}
  where $\cdots$ denotes the omission of the non-linear terms of the
 fields. Therefore, each fermionic field is transformed as
  \begin{eqnarray}
  & & [\delta_{\epsilon}, \delta_{\xi}] \chi (x) = 0 + \cdots, \nonumber \\ 
  & & [\delta_{\epsilon}, \delta_{\xi}] \chi^{(i_{1} \cdots i_{l+1})} (x)
  = - 2 a_{4} (  {\bar \xi}_{1} \Gamma^{j} \epsilon_{1}
      + {\bar \xi}_{2} \Gamma^{j} \epsilon_{2} )
     \chi^{( \{i_{1} \cdots i_{l})} (x) \delta^{i_{l+1} \} j} +
  \cdots. \label{AZ3fermtr}
  \end{eqnarray}
  The commutator of the SUSY transformation constitutes a 
  translation for the vector fields, while the same does not hold true  of 
  the fermionic fields.
  However, the analysis (\ref{AZ3comm-m}) $-$ (\ref{AZ3fermtr}) is
  applicable only to the case in which we can perform a Taylor
  expansion of $V_{S}(u)$ around $u=0$. We
  speculate that we may be able to construct a well-defined 
  model with ${\cal N} =2$ SUSY for the functions like
  (\ref{AZ22example1}), which do not have the Taylor expansion around
  $u=0$.  

  We  surmise that the fermionic
  fields  $\chi(x)$ and $\chi^{(i)} (x)$, which are identified with the
  dilatino and the gravitino respectively, become massless  due to the
  supersymmetry, when the even-rank fields ${a^{(i)}}_{i a_{1} \cdots
  a_{2n}} (x) $ are massless.
  If we choose the function $V_{S} (m^{2})$ properly,
  the classical low-energy limit of this matrix model would reduce to
  the type IIB supergravity, and it is an interesting future problem
  to investigate this conjecture more deeply. 

 \section{Conclusion}
 We have hitherto investigated the possibility to realize the
 general coordinate invariance manifestly in a matrix model. And
 in order to build a matrix model which describes the gravity, we should
 consider the local Lorentz invariance of the model. In realizing
 the local Lorentz invariance in a matrix model, the parameters of the 
 Lorentz symmetry must be promoted to $u(N)$ matrices,
 because the information of the spacetime is encoded 
 in the eigenvalues of the large $N$ matrices. The model
 thus has to include all the odd-rank gamma matrices in 10 dimensions.
 The starting point to 
 construct such a matrix model is to identify the large $N$ matrices
 with differential operators. By this identification, it is
 possible to describe the differential operators on an arbitrary spin
 bundle over an arbitrary manifold simultaneously in the space of
 large $N$ matrices. 

  We have first considered the bosonic part of the model, and
  clarified that it reduces to the Einstein gravity in the classical
  low-energy limit. And we have extended the observation to the
  supersymmetric action, and have presented a
  proposal to construct a matrix model which has ${\cal N}=2$ SUSY and
  reduces to the type IIB supergravity. It is a future problem to seek 
  the explicit form of the function $V_{S}(m^{2})$.
  If we find a proper model, this would be  
  a nonperturbative formulation of the type IIB superstring theory, and
  become an extension of IIB matrix model with the information of the
  gravity encoded more manifestly. It is also interesting to elucidate 
  the relation between our model and IIB matrix model. More
  analysis of these issues will be reported elsewhere. 

 \paragraph{Acknowledgment} \hspace{0mm} \\
 The authors would like to express their gratitude to Keisuke Ohashi for
 valuable discussion.  The works of T.A. were supported in part by
 Grant-in-Aid for Scientific Research from Ministry of Education,
 Culture, Sports, Science and Technology of Japan ($\#$01282). 

\appendix
\section{Examples of the differential operators in the space of 
 large $N$ matrices} \label{diffmatrix}
 This appendix is devoted to some simple examples of the differential
 operators embedded in the space of large $N$ matrices. The
 space of the large $N$ matrices includes the differential operators on
 an arbitrary spin bundle over an arbitrary manifold simultaneously.
 We now  consider the differential operators of the scalar field 
 on two different bundles over $S_{1}$. 
    \begin{figure}[htbp]
   \begin{center}
    \scalebox{.7}{\includegraphics{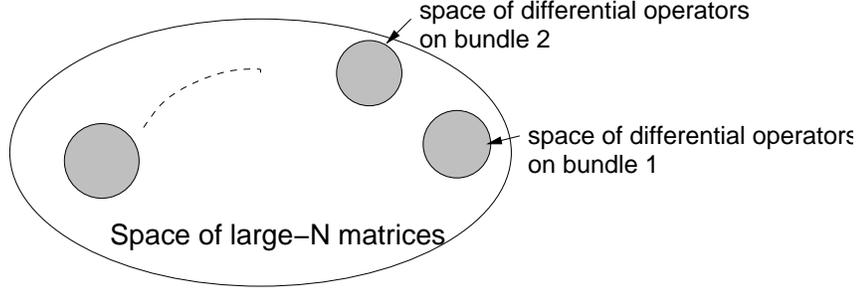} }
   \end{center}
    \caption{The way the spaces of the differential operators are
      embedded in the space of large $N$ matrices.}
   \label{diff-matrix}
  \end{figure}
    \begin{figure}[htbp]
   \begin{center}
    \scalebox{.7}{\includegraphics{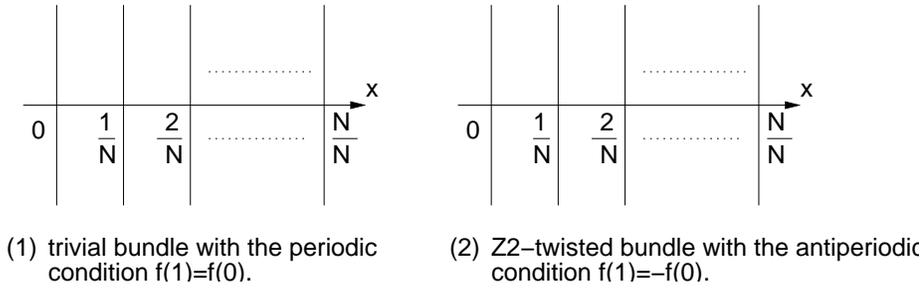} }
   \end{center}
    \caption{Differential operators on (1)the trivial bundle and
      (2)the $Z_{2}$-twisted bundle over $S_{1}$.}  
   \label{diff-matrixex}
  \end{figure}
 \paragraph{(1)trivial bundle} \hspace{0mm} \\
 We first consider the trivial bundle with the periodic condition
 $f(1) = f(0)$. We discritize the region $0 \leq x \leq 1$ into small
 slices of spacing $\epsilon = \frac{1}{N}$.
 Then, the differential operator is approximated by the finite
 difference as 
  \begin{eqnarray}
   \partial_{x} f \left( \frac{k}{N} \right) \to \frac{1}{2} \left( 
   \frac{f (\frac{k+1}{N}) - f(\frac{k}{N})}{\epsilon}  
 + \frac{f(\frac{k}{N}) - f(\frac{k-1}{N})}{\epsilon} \right)
 = \frac{N}{2} \left( f \left( \frac{k+1}{N} \right) -  
                      f \left( \frac{k-1}{N} \right) \right).
  \end{eqnarray}
  Due to the periodic condition, this finite difference is expressed
  by the large $N$ matrix as
  \begin{eqnarray}
  \partial_{x} \to  A = \frac{N}{2} \left( \begin{array}{cccccc}
  0  &  1 &   &   &   &  -1 \\
  -1 &  0 & 1 &   &   &     \\
     & -1 & 0 & 1 &   &     \\
     &    &   & \ddots &  &   \\
  1  &    &   &   & -1 & 0 
   \end{array} \right).
     \end{eqnarray}

 \paragraph{(2)$Z_{2}$-twisted bundle (M{\"o}bius strip)} \hspace{0mm} \\
 We consider the similar problem with respect to the $Z_{2}$-twisted
 bundle, in which the antiperiodic condition $f(1) = - f(0)$ is
 imposed. Paying attention to this antiperiodicity, we understand that 
 the finite difference in the discritized space is expressed as 
  \begin{eqnarray}
  \partial_{x} \to  A = \frac{N}{2} \left( \begin{array}{cccccc}
  0  &  1 &   &   &   & 1 \\
  -1 &  0 & 1 &   &   &     \\
     & -1 & 0 & 1 &   &     \\
     &    &   & \ddots &  &   \\
  - 1  &    &   &   & -1 & 0 
   \end{array} \right).
  \end{eqnarray}

\section{Proof of (\ref{AZ3comm-m}) and (\ref{AZ3comm-psi})}
  \label{AZexplicit3} 
 This appendix is devoted to the explicit computation of the
 commutation relations of the SUSY transformation. 
 First, (\ref{AZ3comm-m})  is verified by considering the differences
 of the following two SUSY transformations: 
  \begin{eqnarray}
   & & m \stackrel{\delta_{\xi}}{\to} m + \xi {\bar \psi} + \psi {\bar 
   \xi}
   \stackrel{\delta_{\epsilon}}{\to} m + (\epsilon + \xi) {\bar \psi}
   + \psi ({\bar \epsilon} + {\bar \xi} ) + 2 \xi {\bar \epsilon}
   V'_{S}(m^{2}) + 2 V'_{S}(m^{2}) \epsilon {\bar \xi}, \nonumber \\
  & & m  \stackrel{\delta_{\epsilon}}{\to} m + \epsilon {\bar \psi} +
   \psi {\bar \epsilon} 
   \stackrel{\delta_{\xi}}{\to} m + (\epsilon + \xi) {\bar \psi}
   + \psi ({\bar \epsilon} + {\bar \xi} ) + 2 \epsilon {\bar \xi}
   V'_{S}(m^{2}) + 2 V'_{S}(m^{2}) \xi {\bar \epsilon}. \nonumber
  \end{eqnarray}
 
 (\ref{AZ3comm-psi}) is verified likewise, but we need to use  the
 Taylor  expansion of $V_{S}(u)$ around $u=0$: 
 {\small
  \begin{eqnarray}
  \psi  &\stackrel{\delta_{\xi}}{\to}& \psi + 2  V'_{S}(m^{2})
  \xi \nonumber \\ 
        &\stackrel{\delta_{\epsilon}}{\to}&
  \psi + 2 V'_{S}(m^{2}) (\epsilon + \psi) 
       + \sum_{k=2}^{\infty} a_{2k} [ (\epsilon {\bar \psi} + \psi
        {\bar \epsilon}) m^{2k-3} + m  (\epsilon {\bar \psi} + \psi
        {\bar \epsilon}) m^{2k-4} + \cdots + m^{2k-3} (\epsilon {\bar
        \psi} + \psi {\bar \epsilon}) ] \xi, \nonumber \\
  \psi  &\stackrel{\delta_{\epsilon}}{\to}& \psi + 2  V'_{S}(m^{2})
  \epsilon \nonumber \\ 
        &\stackrel{\delta_{\xi}}{\to}&
  \psi + 2 V'_{S}(m^{2}) (\epsilon + \psi) 
       + \sum_{k=2}^{\infty} a_{2k} [ (\xi {\bar \psi} + \psi
        {\bar \xi}) m^{2k-3} + m  (\xi {\bar \psi} + \psi
        {\bar \xi}) m^{2k-4} + \cdots + m^{2k-3} (\xi {\bar
        \psi} + \psi {\bar \xi}) ] \epsilon. \nonumber 
   \end{eqnarray} }
  Utilizing the equations of motion (\ref{AZ3EOM}) and the fact
  that the combinations of the fermionic fields ${\bar \psi} \epsilon$
  and ${\bar \psi} \xi$ vanish because the fermions are Weyl ones, we
  find that the commutator of the SUSY transformation is
  \begin{eqnarray}
   [ \delta_{\epsilon}, \delta_{\xi}] \psi 
 =  \sum_{k=2}^{\infty} a_{2k} (\psi {\bar \epsilon} m^{2k-3} \xi
      - \psi {\bar \xi} m^{2k-3} \epsilon) 
 = 2 \psi  \left( {\bar \epsilon} m \frac{V_{S}'(m^{2}) - V_{S}'(0)}{m^{2}}
  \xi - {\bar \xi} m \frac{V_{S}'(m^{2}) - V_{S}'(0)}{m^{2}} \epsilon
  \right). \nonumber  
  \end{eqnarray}
 

 \end{document}